\documentclass[pre,aps,twocolumn,showpacs]{revtex4}
\usepackage{amsmath,amssymb,graphicx}
\usepackage{epsfig}
\usepackage{textcomp}

\begin{document}
\title{Emergence of fluctuating traveling front solutions in macroscopic theory of noisy invasion fronts}
\author{Baruch Meerson}
\email{meerson@cc.huji.ac.il}
\affiliation{Racah Institute of Physics, Hebrew University of
Jerusalem, Jerusalem 91904, Israel}
\author{Pavel V. Sasorov}
\email{pavel.sasorov@gmail.com}
\affiliation{Keldysh Institute of Applied Mathematics, Moscow 125047, Russia}
\author{Arkady Vilenkin}
\email{vilenkin@vms.huji.ac.il}
\affiliation{Racah Institute of Physics, Hebrew University of
Jerusalem, Jerusalem 91904, Israel}

\begin{abstract}
The position of an invasion front, propagating into an unstable state, fluctuates because of the shot noise
coming from the discreteness of reacting particles and stochastic character of the reactions and diffusion.
A recent macroscopic theory [Meerson and Sasorov, Phys. Rev. E \textbf{84}, 030101(R) (2011)] yields the probability of observing, during a long time,  an unusually slow front. The theory
is formulated as an effective Hamiltonian mechanics which operates with the density field and the conjugate  ``momentum" field.
Further, the theory assumes that the  most probable density field history of an unusually slow front represents, up to small corrections, a traveling front solution of the Hamilton equations. Here we verify this assumption by solving the Hamilton equations numerically for models belonging to the directed percolation universality class.
\end{abstract}
\maketitle

\section{Introduction}

The Fisher-Kolmogorov-Petrovsky-Piscounov (FKPP)
equation \cite{Fisher},
\begin{equation}\label{FKPP}
    \partial_t q = q-q^2+\partial_{x}^2 q\,,
\end{equation}
describes invasion of an unstable state, $q(x\to \infty,t)=0$, by a stable state, $q(x\to -\infty,t)=1$.
This equation serves as a fundamental model in mathematical genetics and population biology \cite{Fisher,Murray}.  Closely related
equations appear
in chemical kinetics \cite{Douglas},
extreme value statistics \cite{Majumdar}, dynamics of disordered systems \cite{Spohn}
and even particle physics \cite{particlephys}.

Invasion fronts correspond to traveling front solutions (TFSs) of Eq.~(\ref{FKPP}): $q(x,t)=Q_{0,c}(\xi)$, where $\xi=x-ct$. $Q_{0,c}(\xi)$ solves the  ordinary differential equation
\begin{equation}\label{MFeq}
    Q_{0,c}^{\prime\prime}+c Q_{0,c}^{\prime}+Q_{0,c}-Q_{0,c}^2=0\,,
\end{equation}
where the primes stand for the derivative with respect to the (single) argument.
For a sufficiently steep initial condition, the solution of Eq.~(\ref{FKPP}) approaches at long times the limiting TFS of Eq.~(\ref{MFeq}), $Q_{0,2}(\xi)$, with the velocity $c_0=2$, see Ref. \cite{Saarloos} for a comprehensive review. This special value of the front velocity is determined by the dynamics of the leading edge of the front, where one can linearize Eq.~(\ref{FKPP}) around $q=0$. In other words, the nonlinear front solution, as described by Eq.~(\ref{FKPP}), is ``pulled" by its leading edge, hence the term ``pulled fronts" \cite{Saarloos}, of which the FKPP equation (\ref{FKPP}) is the best studied example.

As a mean-field equation, Eq.~(\ref{FKPP}) does not account for the shot noise coming from the discreteness of particles and stochastic character of the particle reactions and random walk. Even when this noise is week, it causes the front position to strongly deviate from the mean-field theory prediction. This deviation has a systematic part  -- the front velocity shift -- and a fluctuating part. If $N\gg 1$ is the effective number of particles in the front region, the shifted front velocity is \cite{Derridashift,Levine,Derridanum,Derridatheory}
\begin{equation}\label{shiftedc}
c_{*}=2-\frac{\pi^2} {\ln^{2} N}+\frac{6 \pi^2 \ln(\ln N)}{\ln^3 N}+ \dots .
\end{equation}
In its turn, the front diffusion coefficient scales as $\ln^{-3} N$ \cite{Derridanum,Derridatheory,Panja1}. These anomalous properties of noisy pulled fronts are very different from the properties of noisy fronts propagating into \emph{metastable} states. In the latter case the front velocity shift and the front diffusion coefficient both scale as $1/N$ and are therefore much smaller \cite{Kessler,Panja2,MSK,KM}.

The front diffusion coefficient is determined by typical, relatively small fluctuations of the front position. What is the probability of large fluctuations? For negative fluctuations (that is, unusually slow fronts) this question was recently addressed in Ref. \cite{MS2011} in the framework of macroscopic fluctuation theory (MFT). This theory can be derived, using WKB approximation, from the master equation for microscopic lattice models that involve on-site reactions among particles and unbiased random walk \cite{MS,EK1}. The MFT can be formulated as a classical Hamiltonian field theory which involves the particle density field $q(x,t)$ and the canonically conjugate ``momentum" field $p(x,t)$. It is similar in spirit to the mathematically rigorous MFT of Bertini, De Sole, Gabrielli, Jona-Lasinio, and Landim \cite{Bertini}, see also Refs. \cite{Tailleur,DG2009b,van,KMS}, developed for diffusive lattice gases in the absence of on-site reactions. It is also analogous to the Martin-Siggia-Rose field-theoretical formalism  \cite{MSR} for continuous stochastic systems.

Let ${\cal P}(c)$ be the probability density that the fluctuating front moves, during a long time $T\gg 1$, with average velocity $c$ that is smaller than $c_*$. Within the MFT framework, the (minus) logarithm of this probability density 
is proportional to the mechanical action along a certain phase trajectory $q(t), p(t)$ of the Hamiltonian system, see below.  For a given average front velocity $c$,  $q(x,t)$ describes
the optimal (most probable) density history of the fluctuating front. Meerson and Sasorov \cite{MS2011} considered the set of reactions  $A\rightleftarrows2A$ and random walk. The crucial conjecture they made (see also Ref. \cite{MSK}) was that, apart from boundary layers at $t=0$ and $t=T$, the optimal trajectory is
a \emph{traveling front solution}: $q(x,t)=q(x-ct)$ and $p(x,t)=p(x-ct)$, of the Hamilton equations for $q$ and $p$. The traveling front ansatz reduces these equations to ordinary differential equations. The (instanton-type) solutions of these equations were found analytically in Ref. \cite{MS2011} for some values of $c$ [most importantly for $c$ close, but not too close, to $c_*$] and numerically otherwise. The resulting
$\ln {\cal P}(c)$ turns out to be proportional to $T$, strongly non-Gaussian with respect to $c$ and rapidly falling with an increase of $\delta c \equiv c_*-c$.  The $N$-dependence
of  the quantity $-T^{-1} \ln {\cal P}(c)$ undergoes a major change depending on $\delta c$:
 \begin{equation}\label{Ndependence}
\!-\frac{\ln {\cal P}(c)}{T}\sim \left\{\begin{array}{ll}
\!\frac{\text{const}}{\ln^3 N}\,\exp\left(\frac{\delta c\ln^3 N}{2\pi^2}\right), \! &\mbox{$\frac{2 \pi^2}
{\ln^3 N}\ll \delta c \ll \frac{\pi^2}{\ln^2 N}$}, \\
\!\text{const}\,N e^{-\frac{\pi}{\sqrt{2-c}}}, &\! \mbox{$\frac{2 \pi^2}{\ln^3 N} \ll \delta c \ll 1$}.
\end{array}
\right.
\end{equation}
The first line of Eq.~(\ref{Ndependence}) follows from the second one when $\delta c\ll \pi^2/\ln^2 N$. Furthermore, the first line coincides (up to the numerical pre-factor which is apparently non-universal) with prediction from the phenomenological theory of Derrida \textit{et al.} \cite{Derridatheory}.  The validity of MFT demands that $\delta c \gg 2\pi^2 \ln^{-3} N$ \cite{MS2011}. Needless to say, $N$ must be huge to have these pronounced asymptotic regimes.

It was observed in Ref. \cite{MS2011} that the aforementioned properties of $\ln {\cal P}(c)$ also hold for all sets of on-site reactions which belong to the directed percolation universality class: under condition that the system is sufficiently close to the characteristic (transcritical) bifurcation of the mean-field theory. This is the setting
we address here. We present strong numerical evidence that the TFS is indeed the true optimal history of the particle density field, and it yields the leading-order asymptotics of $\ln {\cal P}$ as described by the scaling relation (\ref{scaling}).

\section{Model}

Consider three on-site reactions: branching $A\stackrel{\lambda}{\rightarrow} 2A$,  coagulation $2A\stackrel{\sigma}{\rightarrow} A$ and decay $A\stackrel{\mu}{\rightarrow} \emptyset$.  These reactions constitute a Markov birth-death process with the birth rate $\lambda(n_i)=\lambda_0 n_i$ and the death rate $\mu(n_i)=\mu_0 n_i+(\sigma_0/2)  n_i (n_i-1)$, where $n_i$ is the number of particles on site $i$. Let us define $B=\lambda_0/\mu_0$ and $K=2\lambda_0/\sigma_0$, and assume that $K\gg 1$. The deterministic rate equation for the on-site dynamics is
\begin{equation*}
\dot{n}=\mu_0 n (B-1-Bn/K).
\end{equation*}
When $B>1$, $n=0$ is a repelling point, and $n=K(1-1/B)$ is an attracting point.
Adding to these reactions symmetric and independent random walk of the particles between neighboring sites, with rate constant $D_0$, we obtain a spatial model that describes invasion of the unstable state $n=0$ by the stable state $n=K(1-1/B)$. We assume that the system is close to its (transcritical) bifuraction at $B=1$, by putting $B=1+\delta$, where $0<\delta \ll 1$.  Under this assumption the model becomes universal: all models belonging to the directed percolation universality class, see e.g. Ref. \cite{EK2}, behave in the same way when properly rescaled.  We also assume that  the hopping rate is sufficiently high: $D_0\gg \mu_0 \delta$, allowing a continuum description in space. Then the mean-field theory of this system is described by the FKPP equation~(\ref{FKPP}), where $q=n/(K\delta)$, $t$ and $x$ are rescaled as follows: $\mu_0 \delta t \to t$ and $x/\ell \to x$, $\ell = [D_0/(\mu_0 \delta)]^{1/2}$ is the characteristic diffusion length, and the lattice spacing is set to one. In their turn, fluctuations of the front position are encoded in the MFT equations that can be derived in WKB approximation \cite{EK2,MS}:
\begin{eqnarray}
    \partial_t q &=&q-q^2+2qp +\partial_x^2q,\label{qt}\\
    \partial_t p &=& -p-p^2+2 q p - \partial_x^2p. \label{pt}
\end{eqnarray}
Equations (\ref{qt}) and (\ref{pt}) are Hamiltonian, with the Hamiltonian
\begin{equation}\label{w}
    H =\int_{-\infty}^{\infty} dx \,w, \;\;\;\; w=qp (p-q+1) -\partial_x q \,\partial_x p\,.
\end{equation}
If $p(x,t)=0$, Eq.~(\ref{qt}) coincides with the FKPP equation (\ref{FKPP}), whereas Eq.~(\ref{pt}) is obeyed identically.
This is the noiseless solution. Noisy fronts have a non-zero $p(x,t)$ which enables them to move with average velocities less than $c_*$.

The boundary conditions for Eqs.~(\ref{qt}) and (\ref{pt}) in $x$ and $t$ are specified as follows \cite{MS2011}. At $x\to -\infty$ there is a stationary distribution of the particle density, sharply peaked at $q=1$.
Therefore, we demand
\begin{equation}\label{minusinf}
    q(-\infty,t)=1 \;\;\;\text{and}\;\;\;p(-\infty,t)=0,
\end{equation}
which corresponds to the deterministic fixed point $(q=1,p=0)$ of the \emph{on-site} Hamiltonian $H_0(q,p)=qp (p-q+1)$. At $x=\infty$ we demand
\begin{equation}
\label{plusinf}
q(\infty,t)=0,
\end{equation}
whereas $p(x,t)$ must be bounded at finite $x$. The front positions at $t=0$ and $t=T\gg 1$ are specified by kink-like particle density profiles, situated at the
distance $X$ apart; $X$ can be positive, negative or zero. The kinks interpolate monotonically between $q(x=-\infty)=1$ and $q(x=\infty)=0$ and decay sufficiently rapidly at $x\to \infty$.  These boundary conditions specify the problem completely. 

Once the Hamilton equations~(\ref{qt}) and (\ref{pt}) are solved, we can calculate the mechanical action $\mathcal{S}$ along the phase space trajectory $q(x,t),\,p(x,t)$ and evaluate the probability density ${\cal P}(X,T)$:
\begin{eqnarray}
 -\frac{1}{N} \ln {\cal P}(X,T)&\simeq&  \mathcal{S} (X,T) =\int_{-\infty}^{\infty} dx \int_0^{T} dt \, (p \partial_t q-w) \nonumber \\
&=& \int_{-\infty}^{\infty} dx \int_0^{T} dt\, q(x,t)\, p^2(x,t),
\label{action}
\end{eqnarray}
where $N=K \delta^2 \ell$. Notice that the effective number of particles $N$ (which we assume to be much greater than $1$) does not coincide with the characteristic number of particles inside the diffusion length: the latter quantity is equal to
$K \delta \ell$, not $K\delta^2 \ell$.

Importantly, the same MFT equations (\ref{qt}), (\ref{pt}) and (\ref{action}) can be derived from the following Langevin equation for a noisy FKPP front:
\begin{equation}\label{Langevin}
\partial_t q = q-q^2+\partial_{x}^2 q+\sqrt{\frac{2q}{N}}\,\eta(x,t)\,,
\end{equation}
where $\eta(x,t)$ is a Gaussian white noise with zero mean and unit variance.

\section{Fluctuating traveling front conjecture}

The fluctuating traveling front conjecture \cite{MSK,MS2011} assumes that, at $T\to \infty$, ${\cal S}(X,T)$ has the following scaling form: ${\cal S}(X,T) = T \,{\cal F}(X/T)$, so that
\begin{equation}\label{scaling}
 - \ln {\cal P}(X,T) = N T \,{\cal F}(X/T).
\end{equation}
Furthermore, the large-deviation function ${\cal F}(X/T)$ is mostly contributed to by a TFS of Eqs.~(\ref{qt}) and (\ref{pt}): $q=q(x-ct)$ and $p=p(x-ct)$, where $c=X/T$. This TFS solves the coupled ordinary differential equations
\begin{eqnarray}
 q^{\prime\prime}+c q^{\prime}
 + q-q^2+2q p&=&0\,, \label{ODEQ}\\
 p^{\prime\prime}-c p^{\prime}
 +p+p^2-2q p &=&0\,. \label{ODEP}
\end{eqnarray}
subject to the boundary conditions
\begin{equation}
  q(-\infty)=1,\;\;\;p(-\infty)=0, \;\;\;\text{and}\;\;\;q(\infty)=0,
\label{TFSBC}
\end{equation}
whereas $p(\xi)$ must be bounded at finite $\xi$ \cite{conservation}. Similarly to the model studied
in Ref. \cite{MS2011}, there is a symmetry relation among $q$ and $p$ profiles of the TFS:
\begin{equation}\label{symm1}
    p(\xi)=-q(\xi_0-\xi),
\end{equation}
where $\xi_0$ only depends on $c$. In its turn, Eq.~(\ref{action}) reduces to
\begin{equation}
\label{accumrate}
   -\frac{1}{N T} \ln {\cal P}(c)\simeq {\cal F}(c)= \int_{-\infty}^{\infty}  d\xi \, q(\xi)\, p^2(\xi).
\end{equation}
It is straightforward to solve Eqs.~(\ref{ODEQ}) and (\ref{ODEP}) numerically by using a shooting algorithm
\cite{MSK,MS2011}. Instead, we focus here on the important regime of $2-c \ll 1$ where an analytic perturbation theory can be developed \cite{MS2011}. At $2-c \ll 1$ the
action is mostly gathered from the leading edge of the front, $\xi\gg 1$, where $q\ll 1$.
In the (almost deterministic) ``left region" one has $|p|\ll 1$,
whereas in the (scarcely populated) ``right region" one has $q\ll 1$. A closed analytic theory is possible
because there is a joint region where $|p|\ll 1$ and $q\ll 1$ simultaneously. The matching calculations
coincide with those used in Ref. \cite{MS2011}, and we obtain $q(\xi)$ and $p(\xi)$ in terms
of two overlapping asymptotics:
\begin{equation}\label{Qsol}
q(\xi)=\left\{\begin{array}{ll}
Q_{0,c}(\xi)\,,  &\mbox{$\xi_0-\xi \gg 1$}, \\
-Q_{0,c}^{\prime}(\xi_0-\xi)\, e^{c(\xi_0-\xi)-\xi_0}\,, &\mbox{$\xi\gg 1$},
\end{array}
\right.
\end{equation}
and Eq.~(\ref{symm1}) for $p(\xi)$. Here
\begin{equation}\label{xi0}
     \xi_0  = \pi/\sqrt{2-c} +1 + {\cal O}\left(\sqrt{2-c}\right).
\end{equation}

In its turn, the large deviation function (\ref{accumrate}) becomes
\begin{equation}\label{asymplead}
  {\cal F}= \frac{2}{3 e}\, e^{-\frac{\pi}{\sqrt{2-c}}}
\int_{-\infty}^{\infty}\!\!
e^{2 \zeta} Q_{0,2}^3(\zeta)\,d\zeta \simeq 0.0074 \, e^{-\frac{\pi}{\sqrt{2-c}}}.
\end{equation}
where $Q_{0,2}(\xi)$ is fixed by the demand
that  $Q_{0,2}(\xi\gg 1)\simeq A \xi e^{-\xi}$, whereas the $ e^{-\xi}$ term is absent. When $\delta c\equiv c_*-c\ll \pi^2 \ln^{-2} N$,  Eq.~(\ref{asymplead}) yields the first line of Eq.~(\ref{Ndependence}).

How large must be $T$ for the traveling front asymptotic to give a dominant contribution to the action? For $c=X/T$ not close to $2$, the criterion is simply $T\gg 1$.
As $c$ approaches $2$, the criterion  becomes more stringent. Indeed, here the $q$ and $p$ fronts of the TFS are shifted by a large distance $\xi_0\simeq \pi/\sqrt{2-c}\gg 1$. In the large region between the fronts one has $q\ll 1$ and $|p|\ll 1$, and Eqs.~(\ref{qt}) and (\ref{pt}) can be linearized:
\begin{eqnarray}
    \partial_t q &=&q+\partial_x^2q,\label{qtlin}\\
    \partial_t p &=& -p- \partial_x^2p. \label{ptlin}
\end{eqnarray}
Equation~(\ref{qtlin}) can be solved forward in time starting at $t=0$, whereas Eq.~(\ref{ptlin}) can be solved backward in time starting from $t=T$. The corresponding initial conditions $q(x,0)$ and $p(x,T)$ are localized, so their spread is described by the diffusion and anti-diffusion terms in Eqs.~(\ref{qtlin}) and (\ref{ptlin}), respectively. Therefore, the transient time $\tau$ that it takes for the TFS of the complete problem to set in can be estimated from the condition that the diffusion length $\sqrt{\tau}$ is comparable with $\xi_0$. Therefore, for the TFS to give a dominant contribution to the action, we must demand $T\gg \tau$ which leads to
\begin{equation}
\label{establish}
    T\gg \frac{\pi^2}{2-c}.
\end{equation}
Now, the macroscopic theory is valid  when there are many particles at the leading edge of the front, $\xi\simeq \xi_0\simeq \pi/\sqrt{2-c}$. This demand boils down to $c_{*}-c\gg 2\pi^2 \ln^{-3} N$, with $c_{*}$ from Eq.~(\ref{shiftedc}).   At the border of the applicability region of our theory we have
$2-c\simeq \pi^2/\ln^2 N$, and criterion (\ref{establish}) becomes $T\gg \ln^2 N$. Interestingly,  $\ln^2 N$ coincides with the typical relaxation time of fluctuations contributing significantly to the front diffusion in the phenomenological theory of Derrida \textit{et al}. \cite{Derridatheory}.

\begin{figure}[ht]
\includegraphics [width=2.5 in,clip=]{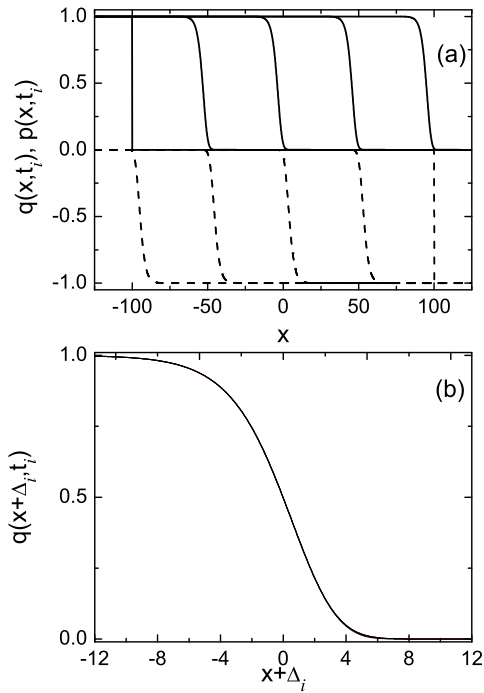}
\caption{(a) The computed density profiles $q(x, t_i)$ (the solid lines) and momentum profiles $p(x, t_i)$ (the dashed lines) at time moments $0, T/4, T/2, 3T/4$ and $T$. The parameters are $X=200$, $T=120$ and $L=250$. (b) The shifted density profiles $q(x+\Delta_{i},t_{i})$ at time moments $T/4, T/2, 3T/4$ and $T$. The shifts $\Delta_i$ are determined from the condition $q(x+\Delta_i=0,t_i)=1/2$. The collapse of the shifted profiles supports the fluctuating traveling front conjecture.}
\label{qp1dt65}
\end{figure}

\begin{figure}[ht]
\includegraphics [width=2.5 in,clip=]{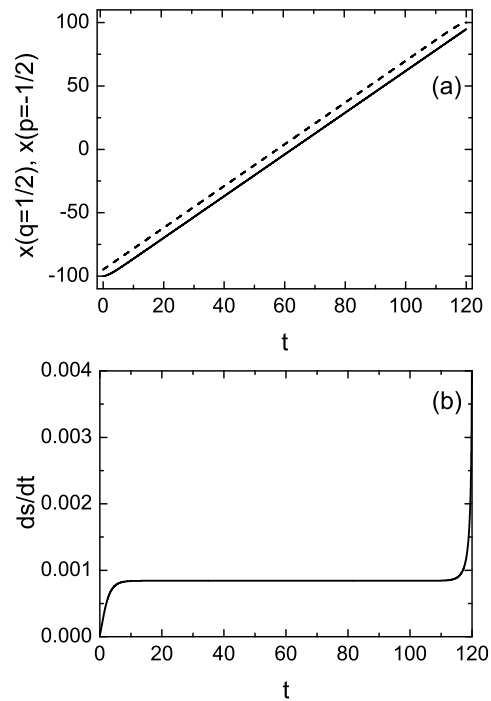}
\caption{(a) The time dependence of the front positions of $q$ (the solid line) and $p$ (the dashed line), defined by the relations $q(x,t)=1/2$ and and $p(x,t)=-1/2$, respectively. The parameters are $X=200$, $T=120$ and $L=250$. The velocity of the traveling front is $c=1.65$ which is close to $X/T=1.66\dots.$ (b) A numerically computed derivative of the accumulated action, $ds/dt$, versus time.}
\label{sqp1dt65}
\end{figure}

\section{Numerics}

To test the predictions of the fluctuating traveling front conjecture, we solved the time-dependent equations
Eqs.~(\ref{qt}) and (\ref{pt}) numerically without any \textit{a priori} assumption about the character of solution.
In particular, the kink-like density profiles at $t=0$ and $t=T$ were quite different  from those
predicted by the TFS. We used an iteration algorithm, originally suggested by Chernykh and Stepanov
\cite{Chernykh} for evaluating the probability density of large negative
velocity gradients in the Burgers turbulence. Different modifications of this algorithm were used for the determination of the optimal history of the density field in several diffusive lattice gas settings, with and without on-site reactions \cite{EK1,MS,KrMe,KMS}. The algorithm iterates the diffusion-type equation~(\ref{qt})
forward in time from $t=0$ to $t=T$, and the anti-diffusion-type equation~(\ref{pt})
backward in time from $t=T$ to $t=0$. Correspondingly, it demands \emph{mixed} boundary conditions in time: $q$ at $t=0$
and $p$ at $t=T$.  This presents an obstacle, as the boundary conditions that we specified involve
the knowledge of $q$ at both $t=0$ and $t=T$. Fortunately, it turns out that the boundary condition for $q(x,T)$ can be replaced by a boundary condition for $p(x,T)$ that has a form of a kink that interpolates monotonically between $p(x=-\infty)=0$ and $p(x=\infty)=-1$. For simplicity we specified
\begin{equation}\label{BCtime}
   q(x,t=0)=\theta(-x), \;\;\; p(x,t=T)=-\theta(X-x),
\end{equation}
where $\theta(x)$ is the Heaviside step function. Note that in this case the momentum field $p(x,t)$ is related to the density field $q(x,t)$ \cite{symmetry} by
\begin{equation}\label{symmetry}
p(x,t)=-q(X-x,T-t),\;\;\;0<t<T.
\end{equation}

We implemented the  Chernykh-Stepanov algorithm using implicit finite differences in iterations of Eqs.~(\ref{qt}) and (\ref{pt}) in a finite box $|x|<L/2$, where  $L>X$. The boundary conditions at $x=\pm\infty$ were replaced by the same conditions at $x=\pm L/2$. Because of the finiteness of the box we also needed a boundary condition for $p(L/2,t)$. We checked that the results in the bulk are insensitive to this condition: $p(x,t)$ always approaches  $-1$ at sufficiently large $x$ and develops a boundary layer at $x=L/2$ to accommodate the specified boundary condition at $x=L/2$. Therefore, we simply put $p(L/2,t)=-1$ in further computations.

We found that the implicit realization of the algorithm is beneficial for convergence of the iterations. This is because the implicit realization guarantees  that $0\leq q(x,t)\leq 1$ and $-1\leq p(x,t)\leq 0$ for all $x$ and $t$, once these two double inequalities hold at $t=0$ and $t=T$, respectively. For the continuous version of the equations, the latter property can be proven as follows.  Let $q>0$ tend to zero at finite $x$. Consider a small vicinity of the minimum point of  $q(x)$ where  $\partial_{x}q=0$ and $\partial^{2}_{x}q>0$.  Here one can neglect all the terms in Eq.~(\ref{qt}) except the diffusion term and see that  $\partial_{t}q>0$ at the minimum point. As a result, $q$ stays positive. Similarly, $q$ cannot reach $1$ at any finite $x$. Indeed, suppose that $q(x,t)<1$ and tends to $1$. Then,  introducing $\epsilon(x,t)=1-q$ and neglecting polynomial terms of order $\epsilon$ and higher, we obtain from Eq.~(\ref{qt}) $\partial_t \epsilon=\partial^{2}_{x}\epsilon-2p$.  Since $p\leq 0$ and  $\partial^{2}_{x}\epsilon>0$ (we consider a vicinity of maximum of $q$ which corresponds to a minimum of $\epsilon$), we have $\partial_t \epsilon > 0$. Therefore, $\epsilon$ cannot reach $0$, and  $q$ cannot reach $1$. In a similar fashion, one can prove that $p(x,t)$ always stays on the interval $(-1,0)$ while evolving backward in time.

The data presented in Figures \ref{qp1dt65} and \ref{sqp1dt65} were computed  for $X=200$, $T=120$
and $L=250$. Figure \ref{qp1dt65}a shows the numerically found spatial profiles of $q(x,t_{i})$ and $p(x,t_{i})$ at different times $t_i$. One can see that, beyond the boundary layers at small $t$ (for $q$) and at $t$ close to $T$ (for $p$), a uniformly translating front develops. This observation is confirmed in Fig. \ref{qp1dt65}b which shows a perfect collapse of shifted profiles $q(x+\Delta_i,t_{i})$ into a single profile. The shifts $\Delta_i$ were determined from the condition $q(x+\Delta_i=0,t_{i})=1/2$. Figure \ref{sqp1dt65}a depicts the positions of the $q$- and $p$-fronts  (defined as the points where $q=1/2$ and $p=-1/2$) versus time. Both of them are straight lines with the same slope. The resulting traveling front velocity $c=1.65$ is close to the expected value $X/T=1.66\dots$.

A sharp signature of a fluctuating traveling front is a linear $t$-dependence of the accumulated action
\begin{equation}
s (t,X,T) = \int_{-\infty}^{\infty} dx \int_0^{t} dt^{\prime}\, q(x,t^{\prime})\, p^2(x,t^{\prime}).
\label{accumaction}
\end{equation}
This linear time dependence is verified in
Fig. \ref{sqp1dt65}b which shows the time derivative of $s(t)$ evaluated numerically. As one can see, this time derivative is constant up to transients at times close to $0$ and to $T$.

\begin{figure}
\includegraphics [width=2.5 in,clip=]{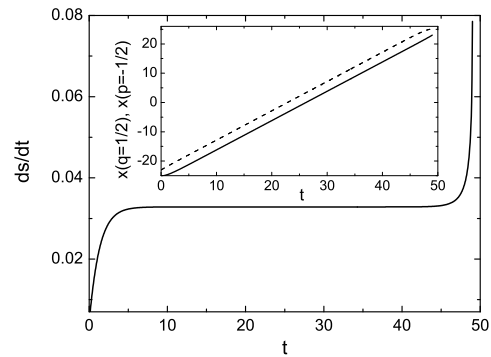}
\caption{A numerically computed derivative of the accumulated action, $ds/dt$, versus time, for $X=50$, $T=49$ and $L=100$.  Inset: the time dependence of the front positions of $q$ (the solid line) and $p$ (the dashed line). The traveling front velocity is $c=1.0$ which is close to $X/T \simeq 1.02$.}
\label{C1}
\end{figure}

\begin{figure}[ht]
\includegraphics [width=2.5 in,clip=]{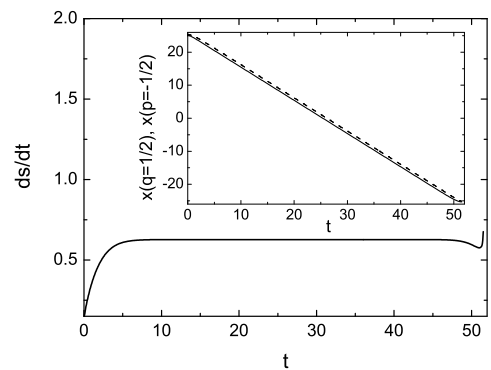}
\caption{Same as in Fig. \ref{C1} but for $X=-50$, $T=51$ and $L=100$.  The traveling front velocity is $c=-1.0$ which is close to $X/T \simeq - 0.98$.}
\label{Cmns1}
\end{figure}

We also observed fluctuating TFSs for other sets of parameters. Two additional examples are shown in Figs.  \ref{C1} and \ref{Cmns1}. Again,  depicted are
the front positions of $q$ and $p$ versus time and the the time derivative of the accumulated action $s$. Constant front velocities ($c=1.0$ and $c=-1.0$ in the respective cases) and constancy of $ds/dt$ versus time are clearly seen, confirming the fluctuating traveling front conjecture. We also checked that the results are insensitive to the exact form of the $q$- and $p$-kinks  at $t=0$ and $t=T$ respectively: under condition that the kinks are well localized. Importantly, in all numerical examples we did observe a kink-like density profile at $t=T$, thus validating \textit{a posteriori} the replacement of the kink-like boundary condition for $q(x,T)$ by a kink-like boundary condition for $p(x,T)$, see Eq.~(\ref{BCtime}).

\begin{figure}[ht]
\includegraphics [width=2.5 in,clip=]{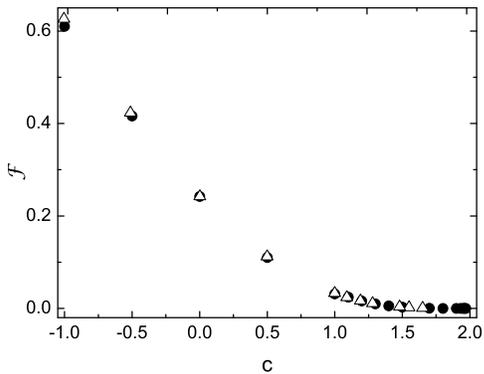}
\caption{Triangles:  the large deviation function ${\cal F}(c)$ corresponding to the plateau region of $ds/dt$  obtained from the time-dependent solutions. Circles: ${\cal F}(c)$ from Eq.~(\ref{accumrate}), obtained by numerically solving the traveling front equations (\ref{ODEQ}) and (\ref{ODEP}).}
\label{Cdsdt}
\end{figure}

Figure \ref{Cdsdt} shows the large deviation function ${\cal F}$, corresponding to the plateau region of $ds/dt$ observed in the time-dependent solutions, versus $c$. Also shown is ${\cal F}$ from Eq.~(\ref{accumrate}), obtained by numerically solving, by a shooting method, the traveling front equations (\ref{ODEQ}) and (\ref{ODEP}). (See Refs. \cite{MSK,MS2011} for details of the shooting method.) One can see very good agreement which again confirms the traveling front conjecture.  When $c$ approaches $2$, ${\cal F}$ is exceedingly small, see Eq.~(\ref{asymplead}), and the accuracy of our time-dependent solution becomes insufficient to probe this regime.

It is crucial that ${\cal F}(c)$ is a convex function, see Fig.~\ref{Cdsdt}.  As a result, at given $X/T$, a TFS with $c=X/T$ has a lesser action (and, therefore, a higher probability) than a front that first moves slower and then faster, or vice versa.

\section{Summary}

Our numerical solution of the time-dependent equations of the macroscopic
fluctuation theory (MFT) of pulled noisy fronts gives a full support to the fluctuating traveling front conjecture of Ref. \cite{MS2011}. We always observed
that, except at $t$ close to $0$ and $T$, the optimal path of the system, as described by the MFT equations, has the form of a traveling front of $q$ and $p$. We also confirmed the scaling behavior~(\ref{scaling}) of the probability to observe an unusually slow front. These findings put the macroscopic theory of unusually slow pulled noisy fronts \cite{MS2011}  on a solid foundation. A major challenge is to develop a theory of unusually \emph{fast} pulled fronts that would go beyond the successful phenomenological theory of Derrida \textit{et al}. \cite{Derridatheory}.

\section*{ACKNOWLEDGMENTS}

B.M. and P.S. are very grateful to Bernard Derrida for discussions and advice. B.M. and A.V. were supported by the Israel Science Foundation (Grant No. 408/08). P.V.S. was supported by  the Russian Foundation for Basic Research, grant No 10-01-00463.


\begin{thebibliography}{99}
\bibitem{conservation} Equations~(\ref{ODEQ}) and (\ref{ODEP}) possess a conservation law: $H_0[q(\xi),p(\xi)]+q^{\prime} p^{\prime}=\mbox{const}$, where the constant is zero because of the boundary conditions.
\end{thebibliography}
\end{document}